# *A single inverse-designed photonic structure that performs parallel computing*

Miguel Camacho, Brian Edwards, Nader Engheta*

University of Pennsylvania
Department of Electrical and Systems Engineering
Philadelphia, PA 19104, USA.
*Correspondence to: engheta@ee.upenn.edu

**In the search for improved computational capabilities, conventional microelectronic computers are facing various problems arising from the miniaturization and concentration of active electronics devices[1,2]. Therefore, researchers have been exploring several paths for the next generation of computing platforms, which could exploit various physical phenomena for solving mathematical problems at higher speeds and larger capacities. Among others, physical systems described by waves, such as photonic and quantum devices, have been utilized to compute the solution of mathematical problems[1–18]. However, previous devices have not fully exploited the linearity of the wave equation, which as we show here, allows for the simultaneous parallel solution of several independent mathematical problems within the same device. In this Letter, we demonstrate, theoretically and experimentally, that a transmissive cavity filled with a judiciously tailored dielectric distribution and embedded in a multi-frequency feedback loop can calculate the solutions of an arbitrary number of mathematical problems simultaneously. We design, build, and test a computing structure at microwave frequencies that solves two independent integral equations with any two arbitrary inputs. We offer another design that can invert four arbitrary 5x5 matrices, confirming its functionality with numerical simulations. We believe our results presented here can provide "coincident computing" and pave the way for the design of low-power, ultrafast, parallel photonic analog computing devices for sensing and signal processing in embedded computing applications.**

From nonlinear image processing[19] to cryptography[20], the ability to quickly solve mathematical problems involving linear algebra is the key for handling the increasing volume of information generated every day. As traditional electronic computers find it more difficult to increase their capabilities by achieving further miniaturization[2], novel approaches based on different platforms have become closer to being functional. Researchers have managed to exploit the laws of physics in such a way that the response of well-known physical systems can be utilized to solve more abstract mathematical problems[21]. Among the most promising scenarios are those based on wave phenomena, such as quantum[6,11,22–24] or guided wave systems[1,14,15,25–28].

While earlier efforts using optical signal processing required extensive free-space propagation to do mathematical operations in a Fourier-transformed space[29–32], more recent approaches have benefitted from the use of compact inverse-designed structures that can reliably reproduce the desired optical functionality in a more limited footprint[7,33–35]. Inverse-design techniques have been largely exploited for the design of complex nanophotonic structures, such as couplers for multiplexing,[36] devices involving non-linear phenomena[37,38], and devices for sound classification[34]. Similar to other photonic devices, in the linear regime these structures can be characterized by a *scattering matrix* connecting a set of complex-valued input modes to a set of complex-valued output modes, thus capturing the operation of matrix multiplication[39]. Other researchers have exploited known resonant geometries to achieve differential equation solvers for time-envelope signals[40], and an example of parallel photonic differentiation has also been recently proposed[41].

By implementing feedback paths into linear systems the output and input modes can be linked together, resulting in platforms that, in addition to performing mathematical operations, can also compute the solution of mathematical problems requiring matrix inversion[7].

In this Letter we demonstrate the ability of wave-based computing platforms to handle and solve several mathematical problems *simultaneously*, each encoded on a wave with a different frequency. These structures, which follow the scheme shown in Fig. 1, are capable of solving equations given by

$$g(u) = e(u) + \int_a^b K(u,v)g(v)dv \qquad (1)$$

which constitutes a Fredholm integral equation of the second kind. In a physical realization of the device based on *N* input and *N* output waveguides, this equation can be discretized into matrix form, $\boldsymbol{g} = \boldsymbol{e} + \boldsymbol{Kg}$, where $\boldsymbol{g}$ represents a $N \times 1$ vectorial solution obtained as the complex-valued signal at each waveguide, $\boldsymbol{e}$ is the $N \times 1$ vector input signal and $\boldsymbol{K}$ is an $N \times N$ discretized version of the kernel $K(u, v)$. The metamaterial device gives the output $\boldsymbol{g} = \text{inv}(\boldsymbol{I}_{nxn} - \boldsymbol{K})\boldsymbol{e}$, where $\boldsymbol{I}_{nxn}$ represents the identity matrix of the same order as the kernel. This mathematical representation provides important insight on the capability of this metamaterial platform, as it allows for the fully analog inversion of a given matrix – an important mathematical process in diverse fields. The problem is then reduced to finding a dielectric distribution whose transmission parameters reproduce the N x N matrix form of the kernel, while having near-zero reflections, where the inputs and outputs are the waveguides on the left and on the right, respectively. In other words, we need to prescribe the transmission part of the S-matrix, $\boldsymbol{T}$, to be $\boldsymbol{K}$.

In an earlier work from our group, this concept was experimentally demonstrated with a reflective cavity which necessitates symmetric kernels[7]. However, in the present work, we embark on a significantly more ambitious approach to fulfill two important goals: (1) To design a single metamaterial structure that can *simultaneously* solve *several* independent equations with different kernels each with arbitrary input signals; and (2) to include more general, i.e., non-symmetric and symmetric, kernels, using the transmissive cavity and feedback mechanism.

To achieve these goals, let us start by considering the following two complex kernels as our first example for two independent Fredholm integral equations of the second kind

$$K_1(u,v) = 0.2 \cos(0.2u) \sin(u-v) + 0.2 \tanh(3v - 0.4iu)\, e^{i(2u-v)} \tag{2}$$

$$K_2(u,v) = 0.2\, J_1(v) + 0.07 \sinh(iv - u)\, e^{i(3u-1)} \tag{3}$$

These two kernels are realized simultaneously in a single metamaterial structure having five inputs and five outputs and dimensions 30 cm x 60 cm shown in Fig.2A. The transmission matrices, $\boldsymbol{T}_1$ and $\boldsymbol{T}_2$, are equal to the two kernels in the interval $u, v \in [-2,2]$ at two different frequencies, i.e. at $f_1$ = 4 GHz for $K_1(u,v)$ and $f_2$ = 5 GHz for $K_2(u,v)$, while requiring negligible reflection. We assume a harmonic time dependence for the fields given by $e^{-i2\pi ft}$.

We utilize the method of inverse design[38] to determine the required permittivity distribution of this structure while enforcing several fabrication constraints: (1) The structure is binary consisting of either air or the commercially available dielectric material Rexolite, with relative permittivity of 2.53 with a loss tangent lower than $10^{-4}$; and (2) The resulting dielectric distribution can be manufactured via standard CNC milling, imposing a minimum internal radius and feature size to allow for fabrication with conventional tooling. Minor post-processing corrections were performed to reinforce these constraints where necessary. The results of this design process are shown in Fig.2A.

In Fig.2B, the resultant design was simulated using commercial software CST Microwave Studio[®42] and the comparison between the simulation results for the complex-valued transmission part of the scattering matrices and the ideal values of the discretized desired kernels at the two frequencies of interest is presented, showing good agreement. The small errors are due to the aforementioned shape post-processing needed for its in-house manufacturing.

Once the planar kernel design is completed, a transmission-based feedback and coupling mechanism is required to excite, recurse, and read the system. While this is theoretically possible in a fully two-dimensional (2D) geometry, a three-dimensional (3D) (i.e., two-level) feedback structure is a suitable way to ensure that the feedback waveguides are of equal length as shown in Fig.2C, thus ensuring equal phase between the feedback lines. Slot couplers are designed to allow total transmission between the upper and lower levels at the two frequencies of interest. To allow for the unidirectional excitation and evaluation of the system, excitation and measurement occur using two pairs of appropriately phased coaxial antennas. Fig. 2D is a photograph of the experimental setup (top off for clarity), showing the dielectric parts fixed in place using plastic pins, the input and output waveguides on either side, and the slot couplers bridging to the hidden waveguides on the backside.

To demonstrate the experimental functionality of our device, we have studied the solutions of our two discretized integral equations with the above selected kernels for five separate excitations, each being a unit-amplitude monochromatic signal at one of the five input ports, which indeed form the basis for any general input to the five input waveguides. These monochromatic signals have one of the two frequencies, $f_1$ and $f_2$ given above, each for one of the two integral equations. The experimentally-extracted solutions at these two frequencies of interest are shown in Fig.2E, compared with those obtained from the full-wave simulation of the complete structure as well as those obtained theoretically by directly inverting the discretized problems, showing an excellent agreement. These results demonstrate the device's ability to simultaneously solve two Fredholm equations of the second kind in a fully analog manner at multiple frequencies.

In the following, as a second example, we theoretically study the extension of our idea to the design of a single metamaterial platform that is able to reproduce four different mathematical kernels simultaneously. In contrast to the previous results, in this case we consider four different matrices, which we would like to invert using our metamaterial device. Here we find the dielectric distribution such that the transmissive part of the S parameters, *T*, is equal to $I - A$, where *A* is the matrix we wish to invert, while the reflection part is required to be negligible. To demonstrate the multi-kernel capability of this approach, four different passive (i.e. that allow for energy conservation) complex-valued matrices have been chosen arbitrarily, $A_1$, $A_2$, $A_3$, and $A_4$. given by:

$$A_1 = \begin{pmatrix} 0.88 + 0.13i & 0.08 - 0.05i & -0.04 - 0.27i & 0.24 + 0.26i & -0.29 - 0.32i \\ 0.22 - 0.31i & 0.65 + 0.16i & -0.16 + 0.28i & 0.1 + 0.09i & 0.2 - 0.2i \\ -0.14 - 0.08i & 0.12 - 0.24i & 0.9 + 0.13i & 0.37 - 0.11i & -0.07 - 0.28i \\ 0.02 - 0.18i & 0.14 - 0.18i & 0.23 + 0.04i & 0.87 + 0.01i & 0.13 - 0.24i \\ -0.25 + 0.23i & 0.19 - 0.27i & -0.26 - 0.19i & -0.04 - 0.07i & 0.86 - 0.2i \end{pmatrix}$$

$$A_2 = \begin{pmatrix} 1.29 - 0.25i & -0.01 + 0.04i & 0.21 - 0.24i & -0.28 + 0.32i & -0.2 - 0.15i \\ -0.04 + 0.11i & 0.97 - 0.15i & -0.08 - 0.1i & 0.41 + 0.21i & -0.11 + 0.01i \\ 0.3 + 0.17i & 0.3 + 0.18i & 0.74 + 0.09i & 0.12 - 0.01i & 0.24 + 0.01i \\ 0.33 + 0.11i & -0.1 - 0.23i & -0.07 - 0.21i & 1.06 - 0.05i & -0.14 + 0.16i \\ -0.01 - 0.04i & -0.29 + 0.14i & -0.3 - 0.31i & -0.26 - 0.04i & 0.67 + 0.22i \end{pmatrix}$$

$$A_3 = \begin{pmatrix} 0.73 + 0.24i & 0.05 - 0.06i & -0.16 - 0.02i & 0.34 - 0.04i & -0.06 - 0.11i \\ 0.17 - 0.28i & 0.79 + 0.2i & -0.05 + 0.24i & -0.2 - 0.07i & 0.2 - 0.13i \\ -0.29 - 0.16i & -0.12 - 0.27i & 1.1 + 0.03i & 0.17 - 0.07i & -0.2 - 0.19i \\ -0.2 - 0.19i & 0.16 + 0.04i & 0.2 + 0.16i & 0.88 + 0.04i & 0.18 - 0.26i \\ -0.16 + 0.17i & 0.31 - 0.26i & 0.03 + 0.08i & 0.14 - 0.02i & 1.13 - 0.27i \end{pmatrix}$$

$$A_4 = \begin{pmatrix} 0.76 + 0.21i & 0.16 - 0.05i & -0.03 - 0.07i & 0.4 + 0.02i & 0 - 0.14i \\ 0.27 - 0.27i & 0.76 + 0.18i & 0.03 + 0.24i & -0.03 - 0.04i & 0.29 - 0.14i \\ -0.15 - 0.14i & 0.03 - 0.25i & 1.04 + 0.05i & 0.3 - 0.07i & -0.07 - 0.2i \\ -0.05 - 0.18i & 0.25 - 0.01i & 0.3 + 0.13i & 0.86 + 0.03i & 0.26 - 0.24i \\ -0.07 + 0.17i & 0.37 - 0.25i & 0.07 + 0.03i & 0.2 - 0.03i & 1.05 - 0.24i \end{pmatrix}$$

Following the previous approach, we select four frequencies within the band of operation of the waveguides, and we apply the optimization method to determine the dielectric distribution for our metamaterial device in order to render the transmission parts of the S parameters associated with the four matrices at four frequencies $f_1 = 4$ GHz, $f_2 = 4.5$ GHz, $f_3 = 5$ GHz, and $f_4$ = 5.5 GHz. The resulting dielectric design is shown in Fig.3A, with the associated transmission matrix, obtained using numerical simulations, shown in Fig.3B. To show the large capability of the presented approach, in this design we imposed that the structure be binary with no other fabrication based constraints, as we did not plan to build this example. We obtained an excellent agreement for all 100 complex transmission coefficients.

Finally, in Fig 3C we present the simulation results from our design showing the inverses of four matrices, compared with the theoretical results expected for these inverses, revealing an excellent agreement.

In conclusion, in this Letter we have demonstrated the capability of single inverse-designed platforms that can simultaneously solve a number of mathematical problems such as integral equations and matrix inversion in a completely independent manner using a multi-frequency approach. An experimental device was constructed and measured at microwave frequencies showing the ability of our approach to solve two independent integral equations simultaneously at two different frequencies involving kernels which are non-symmetric and non-separable, through the use of feedback. The traditionally challenging feedback mechanism was implemented in a compact manner using a multi-level waveguide approach, leading to excellent agreement between the analog solution with respect to the theoretical and simulated solutions. In addition, we have presented a design for a dielectric distribution which could be used for the inversion of four arbitrary matrices using the same approach. We believe that this technology has a strong potential for parallel integrated computing photonic circuitry for fast analog signal processing.

**Acknowledgement**: This work is supported in part by the Vannevar Bush Faculty Fellowship program sponsored by the Basic Research Office of the Assistant Secretary of Defense for Research and Engineering and funded by the Office of Naval Research through grant N00014-16-1-2029.

**Competing interests:** Authors declare no competing interests.

**Data and materials availability:** All data is available in the main text and the supplementary materials.

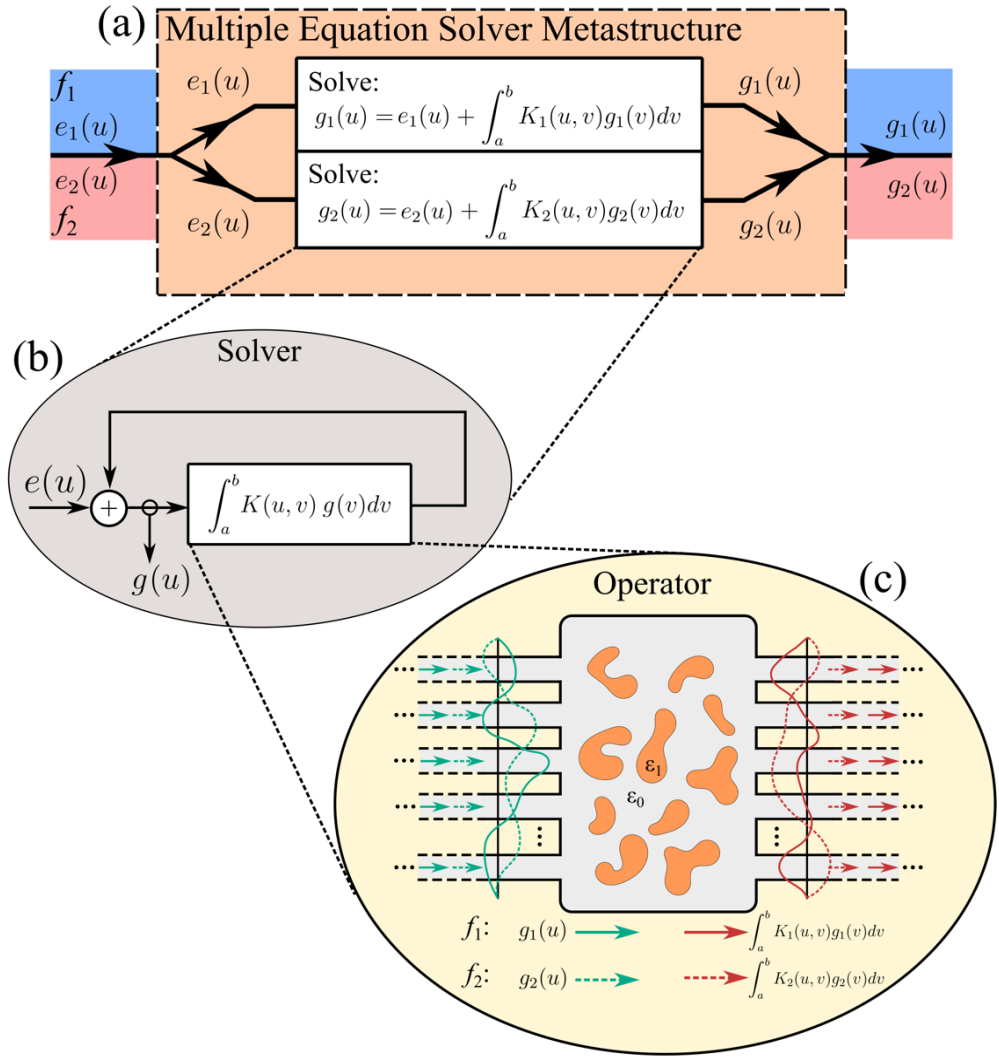

Figure 1: Parallel, material-based, analog computing with waves: The figure shows a conceptual schematic of a device that can simultaneously solve two integral equations as the wave goes through it. This include three levels of design: (a) the device, able to output the solution of two Fredholm integral equations of the second kind encoded in waves at two different frequencies, for which the input waves as the excitation term is introduced into the device, as the arbitrary input to the integral equations. (b) the zero-phase-difference feedback mechanism required for the solution of integral equations. (c) shows the sketch of the physical implementation of the operator block in the form of an inverse-designed dielectric distribution whose transmission matrix corresponds to the values of the kernels of the two Fredholm integral equations of the second kind at the frequencies $f_1$ and $f_2$ respectively.

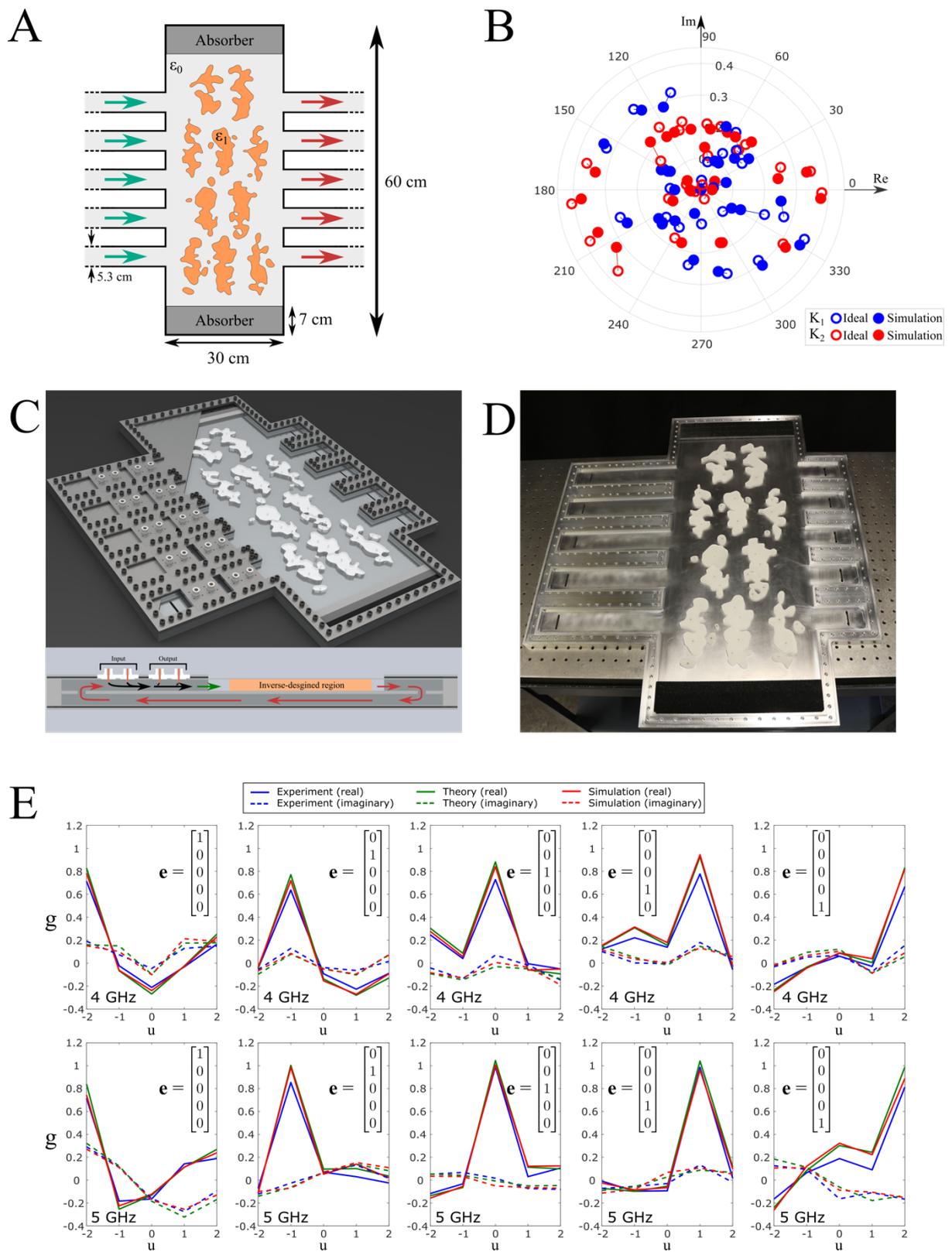

Figure 2: Experimental demonstration of our device that can solve two integral equations at the same time: (A) Inverse-designed dielectric distribution that reproduces the two desired mathematical kernels when two waves with two frequencies $f_1$ and $f_2$ are used. (B) Comparison between the simulation results for the transmission scattering matrices of our design at 4 GHz and 5 GHz, and the corresponding ideal required values for such

*11*

kernels. (C) Rendering of the device including feedback path through a two-level waveguide system, whose schematic is shown in the bottom part. (D) Photograph of the constructed device. (E) Solutions (real and imaginary parts) of the two integral equations for single inputs applied to each of the five input waveguides, obtained theoretically, experimentally and from full-wave simulation.

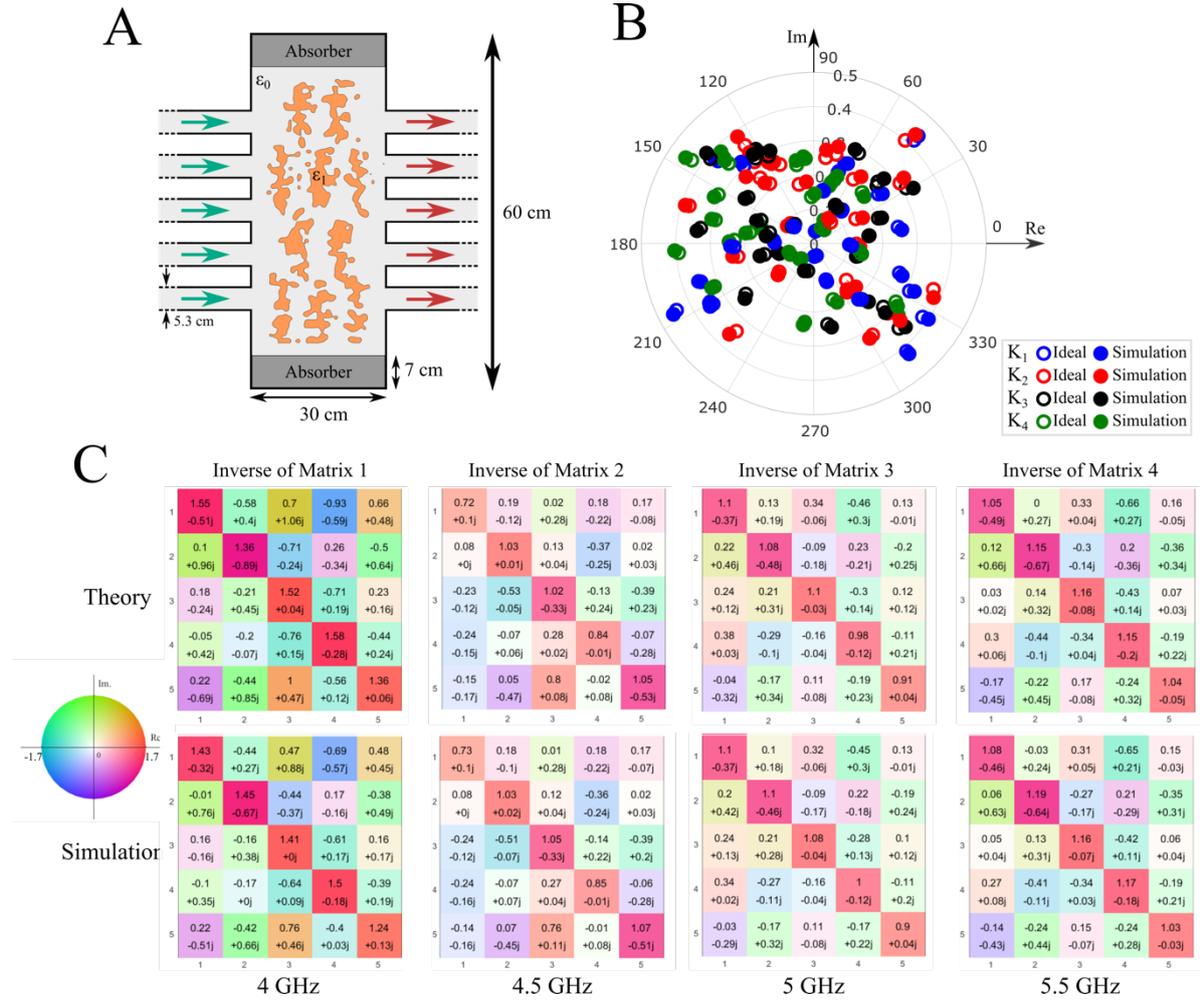

Figure 3: Simulated design for a single metamaterial device that can invert four matrices simultaneously, designed at four different frequencies. (A) Dielectric distribution designed for the four selected matrices. (B) Comparison between the simulation results of our design for the transmission s-parameters of 4 matrices at the four frequencies of interest, and the corresponding desired values of these kernels. (C) Theoretically expected and the simulation results for the inverse matrices achieved using the designed platform.

# *Supplementary Materials for*

**Title:** *A single inverse-designed photonic structure that performs parallel computing*

**Authors:** Miguel Camacho, Brian Edwards, Nader Engheta*

**Affiliations:**

[1]Department of Electrical and Systems Engineering, University of Pennsylvania, Philadelphia, PA 19104, USA.

*Correspondence to: engheta@ee.upenn.edu

**This PDF file includes:**

    Methods

## Methods

*Simulation: Inverse Design and determination of dielectric distribution*

The electromagnetic optimization problem of finding the dielectric distribution associated with the transmission matrix given by the mathematical kernel was undertaken using the topological optimization module commercially available within the finite-element method software COMSOL Multiphysics®. In the software, the objective was set to minimize the complex-plane "distance" between the S-parameters and their desired values, consistent with the transmission part being equal to that obtained from the kernel and the reflection part being zero. A separate "Physics Node" was used for each frequency and for each waveguide input, corresponding to a column in the S-Matrix. Therefore, the number of '"Physics Nodes" required was equal to $N_{wg}N_f$, with $N_{wg}$ being the number of input waveguides and $N_f$ being the number of frequencies for which the design is being optimized. The number of objectives is therefore equal to $2N_{wg}^2 N_f$.

To reduce the family of solutions to those given by binary values of the relative permittivity, a soft step function was introduced as a weighting of the optimization variables when defining the relative permittivity as a function of the former. Several initial conditions were used until a preliminary experimentally-feasible solution was found. The issues arising with radii of curvature and non-binarized points within the distribution were solved by iteratively blurring and reoptimizing the distribution until the issues were resolved. To improve the resolution of the raster-based dielectric distribution before being exported for manufacturing, the mesh size was iteratively reduced and reoptimized until the desired resolution was achieved.

Once the final raster-based design was obtained within COMSOL Multiphysics®, a high-resolution image was exported and processed through an in-house developed contour-analyzer code, which extracted the vector-format shape of each of the closed-polygon structures of the dielectric distribution. This vector-format design was then imported into the three-dimensional (3D) design software SolidWorks® for the generation of the toolpath needed for the milling of the plastic pieces. The vector-based format of SolidWorks® was also used for its simulation in CST Microwave Studio® as a verification step.

*Simulation: Design of the experimental feedback*

The challenge of designing a 3D waveguide structure with feedback was solved by implementing a two-level system. To achieve negligible reflections, a slot was designed to connect both levels, whose dimensions and positions were optimized to give minimal reflections at the frequencies of interest (i.e., 4 GHz and 5 GHz). To accommodate manufacturing limitations, rounded corners (1.59 mm radius) and the metal thickness (63.5 mm) were introduced in the model as provided by the manufacturer. The optimization method used was the trust-region optimization routine implemented in CST Microwave Studio®.

*Simulation: Design of the unidirectional excitation*

The matrix inversion or integral equation solving functionality of the design requires that all waves propagate predominantly only along the direction in which the dielectric distribution is designed to reproduce the desired mathematical kernel. Therefore, to ensure this, we need to enforce that the couplers (which introduce the excitation of the waveguide systems) only excite the mode along the direction of the desired feedback. To do so, we have introduced a pair of coaxial antennas as sources, whose modes are excited with a given phase difference such that the amplitude of the back-propagating mode is zero. The correct phase difference was found through full-wave simulations at the frequencies of interest. To ensure that the presence of these couplers does not perturb the feedback, the length of the tip belonging to the inner conductor of the coaxial antenna has been designed to ensure a coupling coefficient of 5% in amplitude.

*Manufacturing methods*

The system was built using several layers. The main metal structure containing the waveguides on both sides and the cavity was manufactured by an external company (Ephrata Precision Parts Inc.) using industrial CNC milling capabilities. The plastic parts representing the mathematical kernels were manufactured in our lab using a table-top CNC milling machine. Once these pieces and the absorbing material were assembled, both the top and bottom of the system were sealed by laser-cut sheet metal. On top of these, the exciting and probing antennas were held in place using custom-made bridges. Four hundred screws and conductive tapes ensured the electromagnetic isolation of the system, including an additional rubber layer and customized 'washer' on the outside to achieve a uniform force was pressing the top and bottom covers against the sidewall edges. Additionally, the sidewall edges were recessed to ensure that the electrical conduction occurred at the inside corner of the cavity.